\documentclass[aps,floats,twocolumn,preprintnumbers]{revtex4-1}
\usepackage{color}
\usepackage{mathrsfs}
\usepackage{rotating}
\usepackage{graphicx} 
\usepackage{amsfonts}
\usepackage{mathrsfs}
\usepackage[american]{babel}
\usepackage{pstricks}
\usepackage{pst-plot}
\usepackage{pst-grad}
\usepackage{pst-eps}

\newcommand{\bea}{\begin{eqnarray}}
\newcommand{\eea}{\end{eqnarray}}
\newcommand{\beq}{\begin{equation}}
\newcommand{\eeq}{\end{equation}}

\begin{document}
\title{Chiral Phase Transitions around Black Holes}
\author{Antonino Flachi${}^a$ and Takahiro Tanaka${}^b$}
\affiliation{${}^a$Multidisciplinary Center for Astrophysics, Instituto Superior Tecnico, Lisbon, Portugal,\\
%}
%\email{antonino.flachi@ist.utl.pt}
%\author{}
%\affiliation{
${}^b$Yukawa Institute for Theoretical Physics, Kyoto University, Kyoto, Japan}
%\email{tanaka@yukawa.kyoto-u.ac.jp}

%\preprint{preprint-num}
\pacs{pacs}

\begin{abstract}
In this paper we discuss the possibility {that} chiral phase
 transitions, analogous to those of QCD, occur in the vicinity of a
 black hole. If the black hole is surrounded by a gas of strongly
 interacting particles, an inhomogeneous condensate will form. We
 demonstrate this by explicitly constructing {self-consistent} solutions.
\end{abstract}
\maketitle
\vspace{2mm}

%%%%%%%%%%%%%%%%%%%%
%\section{Introduction}
%\label{sec1}

According to the theory of quantum fields in curved space, black holes radiate energy at a temperature inversely proportional to their mass \cite{Hawking:1974sw}. 
As the black hole evaporates, its temperature rises, and at some point a bubble of a high temperature phase surrounding the horizon may form, if a phase transition occurs. This was a particularly interesting phenomena in connection with the Higgs model of electroweak symmetry breaking and its study requires the inclusion of interactions, being an essential feature of the phase transition. 
A method to deal with this 
situation was proposed, but the indication was that, in the Higgs model, the associated high temperature phase would be too localized around the black hole, so that symmetry, effectively, would not be restored \cite{Hawking:1980ng}. The same problem has been reconsidered by Moss taking into account the effect of trapped particles, {\it i.e.} particles emitted by the black hole and reflected back by the walls of the bubble. He indicated that, for some class of bag models, the picture may change and lead to a transient equilibrium configuration of restored symmetry phase, localized around the black hole \cite{Moss:1984zf}. 

A field in which the similar problem of understanding the phase
structure is nowadays very topical is that of QCD at finite temperature
and density, in which phenomena like chiral symmetry breaking and
confinement/deconfinement transitions are known to take place. In this
context, the natural way of addressing the problem would be to use `{\it
first principle}' non-perturbative lattice methods, but already in flat
space, and especially at high densities, things become prohibitive. In
lack of a first principle approach, approximating QCD with a strongly
interacting fermion effective field theories comes in handy. The price
to pay is that {we have} to work with a non-renormalizable effective
theory, but with the bonus of dealing with a simpler one that shares
many of the essential properties of QCD. As a matter of fact, a great
deal of attention is currently {paid on mapping various phases on} the temperature-density diagram within such an effective field theoretical approach,
in order to gain understanding of the vacuum structure of strongly interacting matter (See Ref.~\cite{Fukushima:2010bq} for a recent review). 

The aim of this work is to use the same simplification of degrading QCD to a non-renormalizable, strongly interacting fermion effective field theory, and study the interplay with black holes.
To begin with, we wish to consider a little more in detail the issue of
phase transitions that would break or restore chiral symmetry.
In the context of strongly interacting fermionic systems, it is well
known that chiral symmetry breaking takes place, and this fact is
discussed in terms of the appearance of a fermion condensate. 
{One aspect
particularly important to us is that 
the ground state is believed to develop inhomogeneous phases 
when the density becomes large.} For instance,
Refs.~\cite{Nickel:2009wj} discussed the issue of chiral
symmetry breaking and {the related condensate formation,} and
mapped the phase diagram for models of the Nambu-Jona Lasinio class. 
The description of Refs.~\cite{Nickel:2009wj} 
indicated that the fermion condensate at high densities rensembles a
lattice of domain walls.

In the present case, we are lifting the situation to curved space, where
new effects kick in. In a constant curvature space, the effect of the
non-trivial geometry {is something similar to adding 
chemical potential. The condensate may or may not be spatially
homogeneous. By contrast, in a black hole spacetime 
inhomogeneous configurations for the condensate are inevitable.
To keep the situation as simple as possible, 
let us concentrate on the case in which a Schwarzschild black hole of 
mass $m$ 
is surrounded by strongly interacting fermions in thermal equilibrium 
with the asymptotic temperature given by $T_{BH}=(8\pi m)^{-1}$. 
Then, the local (Tolman) temperature is given by 
$T_{loc}=T_{BH}/\sqrt{f}$ with $f=1-2m/r$.} 
In flat space, the strongly interacting fermionic theory has a critical
temperature, $T_{c}$, that marks {the phase transitions of
chiral symmetry breaking} (in QCD $T_{c} \simeq 200$
MeV). Therefore it seems evident that {in the asymptotic region 
chiral symmetry is restored when $T_{BH}>T_{c}$
while broken for $T_{BH}<T_{c}$. When $T_{BH}<T_{c}$, 
$T_{loc}$ crosses the critical temperature at a certain radius. 
Within this radius, the symmetry will be restored.  
This indicates the possibility that a domain wall structure 
of the condensate surrounding the black hole will arise.} 

We will now make the {above} picture quantitative by using a
strongly interacting fermion effective field theory of the Nambu-Jona
Lasinio type.  
The prototype action can be written as 
\bea
S= \int d^4x \sqrt{g} \left\{ 
\bar \psi i \gamma^\mu \nabla_\mu \psi 
+ {\lambda\over 2N} \left(\bar \psi \psi\right)^2
\right\}~.\nonumber
%\label{action}
\eea
In the above expression $\psi$ is a spinor field, $\lambda$ is the
coupling constant, $g=|\mbox{Det} g_{\mu\nu}|$ is the determinant of the
metric tensor and $\gamma_\mu$ are the gamma matrices in curved
space. 
The number of fermion degrees of freedom (equal to the
number of flavors $\times$ the number of colours) is $N$ and summation over
color and flavor indices is understood. 
The background spacetime is that of a spherically symmetric and
asymptotically flat black hole,
\bea
ds^2&=&fdt^2 +f^{-1}dr^2 +r^2 (d\theta^2 +\sin^2\theta d\varphi^2)~.
\label{schw}
\eea
{The formul\ae~we will present below generally apply for any function $f(r)$}, 
but the numerical analysis will be carried out for the Schwarzschild
case. 

To analyze the breaking/restoration of chiral symmetry, we will use the
finite temperature effective action in the large-$N$ approximation. The
effective action (per fermion degree of freedom), $\Gamma$, can be
expressed, after bosonization, as 
\bea
\Gamma= - \int d^4x \sqrt{g} \left({\sigma^2\over 2\lambda}\right) + 
\mbox{Tr} \ln \left( i \gamma^\mu \nabla_\mu - \sigma \right)~,
\nonumber
\eea
where the composite operator 
$\sigma \equiv-{\lambda\over N} \bar \psi \psi$ was introduced and 
the determinant acts both on field and coordinate spaces. 
Chiral symmetry is broken dynamically when $\sigma$ acquires a non-zero vacuum expectation
value and then a fermion mass term appears. 

The computation of the effective action can be performed using the method
described in Ref.~\cite{Flachi:2010yz}, although some modifications 
are necessary to include the case of black holes. 
{Since black hole spacetimes are static but not 
ultrastatic, we rescale the
metric (\ref{schw}) so as to be ultrastatic, $d\hat{s}^2 =
f^{-1} ds^2$. We will use a hat to indicate the quantities evaluated in
this conformally related spacetime.} After the conformal transformation,
one can use the method of \cite{Flachi:2010yz} to evaluate the effective
action in the rescaled spacetime, $\hat{\Gamma}$, and add a correction
term, $\delta \Gamma$, sometimes called {\it cocycle function}, to
compensate the effect of the conformal transformation~\cite{Dowker:1989gw,Page:1982fm}. 
Assuming the condensate to be spherically symmetric, $\sigma=\sigma(r)$,
and squaring the Dirac operator, 
we obtain 
\bea
\Gamma &=&
-\int d^{4}x \sqrt{g} \left({\sigma^2\over 2\lambda}\right) +\hat{\Gamma}+ \delta \Gamma~,
\eea
where
\bea
\hat{\Gamma}= {1\over 2} \sum_{\epsilon=\pm }\mbox{Tr} \ln \left[
\hat{\square} +\mathscr{A} +f\sigma_\epsilon^2
\right]~.
\label{eff}
\eea
In the above expression $\hat{\square}$ is the D'Alembertian in the
conformally rescaled spacetime and $\sigma_\epsilon^2 := \sigma^2
+\epsilon f^{1/2}\sigma'$. {The quantity 
$\mathscr{A}^{(n)}= f\left((n-2)\Delta \ln f/4-(n-2)^2(\nabla
\ln f)^2/16\right)$ is determined 
so that $\hat{\square}+\mathscr{A}^{(n)}=f^{-(n+2)/4}\square f^{(2-n)/4}$ is 
satisfied, where $n$ is the spacetime dimensions. }
 In (\ref{eff}) we have used the notation
$\mathscr{A}\equiv \mathscr{A}^{(4)}$.
Notice that $\mathscr{A}= \hat{R}/6$. 
{Imposing the periodicity in the Euclidean time with 
the period $\beta=2\pi/T_{BH}$, we express $\hat{\Gamma}$ as}
\bea
\hat{\Gamma}&=& 
{1\over 2} \sum_{\epsilon=\pm }
\sum_{n=-\infty}^\infty
\mbox{Tr} \ln  \left[
-\hat{\Delta} + \omega_n^2 +\mathscr{A} +f\sigma_\epsilon^2
\right],~\nonumber
\eea
with $\hat{\Delta}$ being the Laplacian in the conformally rescaled
space and $\omega_n:={2\pi /\beta}\left(n+1/2\right)$. 

Using zeta regularization gives
\bea
\hat{\Gamma}&=& 
{1\over 2} \int d^{3}x \sqrt{\hat g} \left[ \zeta(0) \ln \ell^2 +\zeta'(0) \right],
\nonumber
%\label{eff2}
\eea
where $\ell$ is a renormalization (length) scale and 
\bea
\zeta(s) := {1\over \Gamma(s)} \sum_{n, \epsilon} \int dt t^{s-1} \mbox{Tr}~e^{-t \left(-\hat{\Delta} + \omega_n^2 +\mathscr{A} +f\sigma_\epsilon^2\right)}.~\nonumber
%\label{zeta}
\eea
The quantities $\zeta(0)$ and $\zeta'(0)$ are the analytically continued
values of $\zeta(s)$ and its derivative to $s=0$. 
The computation of the effective action is rather involved, but it can
be performed in a straightforward manner following the method developed
in Ref.~\cite{Flachi:2010yz}. Here we use a resummed form for the
heat-trace and retain all terms {that contains a specified number of
spatial differentiations.}
We invite the reader to
consult Ref.~\cite{Flachi:2010yz} for details and further references. In
the present case, the result is
%can be written in the following form
\bea
\hat{\Gamma}= 
{{\bf \beta}\over 2(4\pi)^2} 
\sum_\epsilon &\int& d^{3}x \sqrt{\hat g} 
\Bigg\{
{3 \sigma_\epsilon^4\over 4} 
- \left({\sigma_\epsilon^4\over 2} + a_{\epsilon} \right)
\ln \left({f \sigma_\epsilon^2 \over \ell^{2}}\right) 
\nonumber\\
&+&16{\sigma_\epsilon^2 \over f\beta^2}\varpi_2(f^{1\over 2} \sigma_\epsilon) +4 a_{\epsilon} \varpi_0(f^{1\over 2} \sigma_\epsilon)
\Bigg\},~
\label{eff3}
\eea
where we have defined 
\bea
\varpi_\nu (u) &:=& \sum_{n=1}^\infty(-1)^n n^{-\nu} K_\nu\left(n \beta u \right)~,\nonumber\\
a_{\epsilon} &:=& {1\over 180}\left(\hat{R}_{\mu\nu\tau\rho}^2-\hat{R}_{\mu\nu}^2 - \hat\Delta \hat{R}\right)+{1\over 6}\hat{\Delta} \left(f\sigma_\epsilon^2\right)~.\nonumber
\eea
The other term to compute is the cocycle contribution that compensates the difference due to the conformal transformation to recover the result in the original spacetime. The cocycle function can be
expressed in terms of the heat-kernel coefficients associated to the
operator $\mathscr{O}$ in $n$ dimensions:
\bea
\delta\Gamma = \lim_{n\rightarrow 4}\left(C^{(2)}_n[\hat{g}]-C^{(2)}_n[g]\right)/(n-4)~.
\nonumber
\eea
For an operator of the form $\mathscr{O} = \square +V$, 
the part of the heat-kernel coefficient, relevant for our computation, is
\bea
C^{(2)}_n[g] = {1\over (4\pi)^{n\over 2}}{1\over 2} \int d^nx \sqrt{g} \left(V^2 - {1\over 3} R V+ \cdots\right),\nonumber
\eea
where the dots represent terms that do not depend on $V$ or disappear
upon integration by parts. In the present case, 
$V= \sigma_\epsilon^2$ in the original spacetime 
while $V= \mathscr{A}+f\sigma_\epsilon^2$ in the conformally rescaled
spacetime. Simple computations give
\bea
\delta \Gamma &=& {{\bf \beta}\over 2 (4\pi)^{2}}
\sum_{\epsilon=\pm} \int d^3x \sqrt{g}
\left[
{\sigma_\epsilon^4\over 2}\ln f - {2\sigma_\epsilon^2\over f} \lim_{n\rightarrow 4}{d\Lambda_n\over dn}
\right]~,\nonumber
\eea
where $\Lambda_n =\mathscr{A}^{(n)}-(\hat{R}^{(n)}- f R^{(n)})/6$ and
$\hat{R}^{(n)}$ and $R^{(n)}$ are the $n$-dimensional Ricci scalars in
the conformally rescaled and original spacetimes, respectively. Using
$\lim_{n\rightarrow 4}d\Lambda_n/dn = (f^{'2}-2ff''+4ff'/r)/24$, one
arrives at the expression for the cocycle function. 
Combining (\ref{eff3}) with the above expression gives the effective
action $\Gamma$ for the condensate $\sigma$. 

The problem is now reduced to {finding extrema of} the effective
action $\Gamma$ with respect to the condensate $\sigma$. 
Ignoring fourth order derivatives of the condensate allows us to express the equation of motion for the condensate as a 
%second order 
non-linear Schr\"odinger-like equation of the form
\bea
\sigma'' + \delta_1 \sigma' +\delta_2 \sigma'^2 + \mathscr{K}
= 0~,
\label{eq9}
\eea
where the coefficients $\delta_i$ and $\mathscr{K}$ {are functions of
$\sigma$ but independent of its derivatives.}
The explicit expressions are rather long and {will not} be reported here. 

\begin{figure}[ht]
\unitlength=1.1mm
\begin{picture}(70,45)
   \put(-4,18){\textcolor{blue}{\rotatebox{90}{$U_{as}(\sigma)$}}}
   \put(36,-2.5){\textcolor{blue}{$\sigma$}}
   \includegraphics[height=4.8cm]{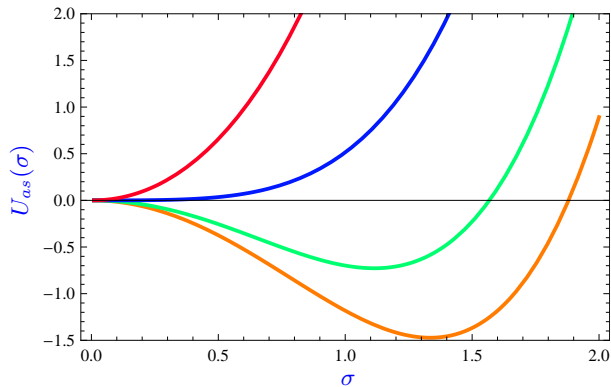}
\end{picture}
\caption{The figure illustrates how, asymptotically, the potential $U_{as}(\sigma)$ changes and symmetry gets restored as temperature increases (The top (red) curve corresponds to $T_{BH}/T_c= 1.75$, while the bottom (orange) curve corresponds to $T_{BH}/T_c = 0.03$. The second curve from top (blue) corresponds to $T/T_c=1$.). We set
%The other parameters are set to 
$\ell=10^{6}$ and $\lambda=10^{-2}$.
\vspace{-0.23cm}} 
\label{fig1}
\end{figure}

Before finding the explicit solution for the condensate, we will
{discuss the critical temperature in the asymptotic region $r\to\infty$. 
Denoting the minus of the action with $\sigma'=0$ as the potential
$U(\sigma)$, the derivative of the asymptotic value can be
computed exactly as }
%\begin{widetext}
\bea
\partial_\sigma {U}_{as}
=-\frac{3 \sigma  \left(4 \lambda \sigma (4 \varpi_{-1}(\sigma)+\beta \sigma  \ln \left(\sigma/\ell\right)-2 \lambda \beta  \sigma^2+\beta \right)}
{2 \lambda \beta (-4 \beta  \sigma  \varpi_{1}(\sigma)-6 \varpi_0(\sigma)+3 \ln\left(\sigma/\ell\right)-2)}.
\nonumber
\eea
%\end{widetext}
The critical temperature is determined by the equation
$\partial_\sigma^2{U_{as}}(\sigma)=0$. Thus, expanding the
Bessel functions contained in $\varpi_\nu$ for small $\sigma$, performing exactly the sum over
$n$, and finally solving a trivial algebraic equation, one arrives at
$T_{c} = \sqrt{3}\lambda^{-1/2}$. {The thermodynamic potential 
obtained by numerically integrating 
$\partial_\sigma {U}$ with respect to $\sigma$ is shown in Fig.~\ref{fig1}}. 

Computing the thermodynamic potential locally will provide further
insight on the form of the condensate. In fact, such a computation shows
that starting from a set of parameters for which asymptotically the
potential has a non vanishing minima, as we move towards the hole, the
minima of the potential will gradually shift towards a configuration
with vanishing $\sigma$.
We confirm the above picture by solving Eq.~(\ref{eq9}) for the
condensate with regular boundary conditions at the horizon. 
Solving Eq.~(\ref{eq9}) can be handled
by standard numerical techniques, but it requires some {caution}. 
First of all, we notice that the coefficients of Eq.~(\ref{eq9}) for $\sigma$
depend on infinite summations over Bessel functions, whose argument is
proportional to the condensate. When the value of the condensate is not
small, these sums can be truncated due to the exponential fall-off of
the Bessel functions. However, when the condensate is small, {fully
resummed expressions have to be used. Once we expand the Bessel functions
for small values of their arguments, we can perform the full resummation
over $n$. 
In the region $r<r_*$ where 
$\sigma$ is small up to a value $r_*$, we integrate Eq.~(\ref{eq9}) 
using this resummed form, and then we switch to the truncated form for
$r>r_*$, matching the value of $\sigma$ and its derivative at the 
junction. 
The boundary conditions in the vicinity of the horizon and 
in the asymptotically far region are set
by requiring that the condensate is at a minima of the potential.} 

\begin{figure}[ht]
%\begin{center}
\unitlength=1.1mm
\begin{picture}(70,45)
   \put(-5,20){\textcolor{black}{\rotatebox{90}{$\sigma^2(r)$}}}
   \put(36,-2.5){\textcolor{black}{$r/r_s$}}
   \includegraphics[height=4.8cm]{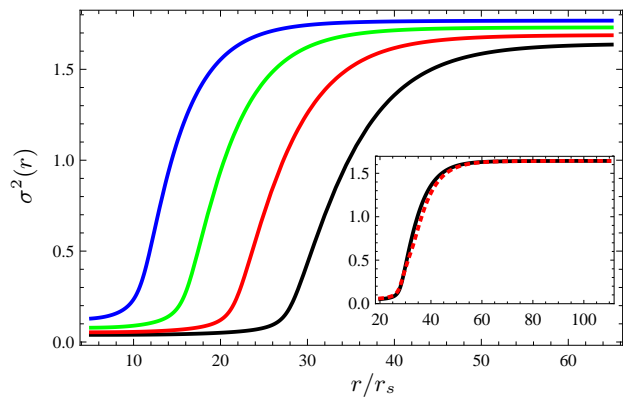}
\end{picture}
%\end{center}
\caption{The figure illustrates the condensate profile found by solving Eq.~(\ref{eq9}), for four indicative values of the black hole temperature (Left to right: $T_{BH}/T_c=0.50~\mbox{(Blue)},~0.54~\mbox{(Green)},~0.58~\mbox{(Red)},~0.61~\mbox{(Black)}$). The values of the other parameters are set to $\ell=10^3$, $\lambda=10^{-2}$. As we increase the black hole temperature, the region of restored symmetry phase expands, and the bubble becomes larger and thicker. The asymptotic value of the condensate becomes smaller as the asymptotic temperatures increases, and tends to zero for $T\rightarrow T_c$. The small box superposed illustrates for the rightmost curve ($T/T_c=0.61$), the corrected solution (red, dashed) when fourth order derivative terms are included.\vspace{-0.2cm}}  
\label{fig2}
\end{figure}
\begin{figure}[ht]
%\begin{center}
\unitlength=1.1mm
\begin{picture}(70,45)
   \put(-5,20){\textcolor{black}{\rotatebox{90}{$\sigma(r)$}}}
   \put(36,-2.5){\textcolor{black}{$r/r_s$}}
   \includegraphics[height=4.8cm]{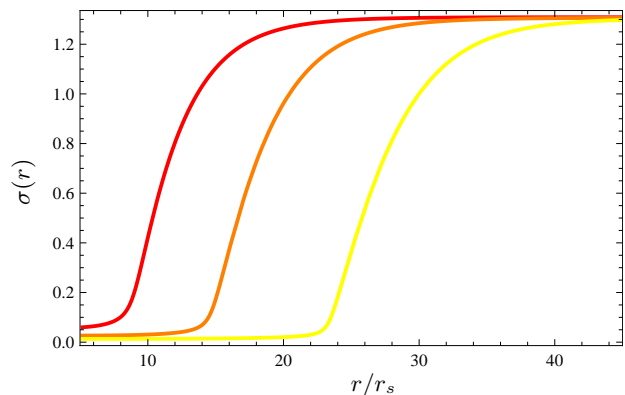}
\end{picture}
%\end{center}
\caption{The figure illustrates the condensate profile for values of the black hole temperature much smaller than $T_c$. The curves refer to (left to right): $T_{BH}/T_c=0.34~\mbox{(Red)},~0.35~\mbox{(Orange)},~0.37~~\mbox{(Yellow)}$.\vspace{-0.5cm}}  
\label{fig2}
\end{figure}
We present the results for the condensate profile in
Fig.~\ref{fig2} for sample values of the parameters. 
The kink-type configurations of Fig.~\ref{fig2} are bubbles that
separate a region of restored symmetry near
the black hole from a region of broken symmetry surrounding it. 
The size of the bubble can also be easily
estimated by equating the local temperature to the critical
temperature as 
\bea
r_{bubble}\sim {r_s}/\left( 1-{T^2_{BH}/T^2_{c}}\right)~,
\nonumber
\label{rbub}
\eea
which approximately agrees with the numerical results.

Higher order corrections can be treated systematically in the present
scheme by {iteratively including higher order} terms in the heat-kernel
expansion. {We expand $\sigma$ around the solution already 
obtained in the lower order approximation, $\bar\sigma$, as 
$\sigma =\bar{\sigma} + \delta \sigma$. Substituting this form 
into the equation of motion, and suppressing all derivatives 
higher than three acting on $\delta\sigma$, we obtain a second 
order differential equation for $\delta\sigma$ with a source term. 
We carried out this iteration to the forth order and verified that 
higher order corrections only produce small distortions 
compared with the solution truncated at the second order. 
In fact, higher order terms become less and less relevant as the black hole temperature
gets closer to the critical one, due to the fact that the 
kink becomes increasingly thicker. In Fig.~\ref{fig2} we have shown
the perturbed solution superposed to the one truncated at second order for an
illustrative set of the parameters. }

The computation presented here considers a situation of 
thermal equilibrium, and thus we used the
Hartle-Hawking vacuum state. To describe evaporating black holes, we
have to use the Unruh vacuum state. The outgoing flux in the 
Unruh vacuum state is diluted at infinity, effectively 
lowering the asymptotic temperature. This suggests that the formation
of a bubble can occur also for black
holes with temperature higher than $T_c$. This case is technically more
challenging and work in this direction is in progress. 

The present discussion may be of 
relevance in the context of primordial black holes. Recent attention has been drawn to
the possibility of a chromosphere formation around the black hole, and
Ref.~\cite{MacGibbon:2007yq}, contrary to the original claim of
Ref.~\cite{Heckler:1995qq}, suggested that this does not happen. 
{We should stress that 
the phase transition across the kink discussed here is a phenomena largely 
different from the chromosphere formation.
For the appearance of the symmetry restored phase around a black hole,  
scattering or reflection on the phase boundary of particles
are not so essential. In the context discussed in this paper, 
the gradient of the effective 
local temperature caused by redshift plays an essential role. 
In the case of the Unruh vacuum, as mentioned above, geometrical dilution 
of particles assists this tendency.  
It would be
definitely interesting to discuss whether the 
appearance of the symmetry restored phase around a black hole   
may change the discussion of
Ref.~\cite{Heckler:1995qq,MacGibbon:2007yq}.}

Several other interesting generalizations of the present work include
the case of charged/rotating black holes. Analyzing the case of higher
dimensional black holes would also be interesting in view of the
possibility that microscopic black holes may form at the LHC. 

Here we considered the simplest class of models and completely ignored
the role of gauge degrees of freedom or other types of condensates, such
as pseudo-scalar ones. Improving the description in this direction is
essential and may help us to gain insight into possible
confinement/deconfinement transitions. Again, the analogy with QCD turns
out to be of use and, in the context of the effective theory approach
adopted here, one natural possibility is to couple the model to the
Polyakov loop \cite{Fukushima:2003fw}. While adding a pseudo-scalar
condensate does not produce {significant changes},
%difficulties} and can be handled technically in the same way, 
the inclusion of gauge fields requires some
additional efforts. Work in this direction is in progress. 

From this point on, speculation would bring us too far, but certainly
there are various interesting problems to consider in relation to the
discussion presented here.

\section*{Acknowledgements}
We wish to thank K. Fukushima, C. Germani, Y. Hikida, M. Nitta,
M. Ruggieri for discussions, comments, and suggestions. The
financial support of the Funda\c{c}\~{a}o p\^{a}ra a  Ciencia e a
Tecnologia of Portugal (FCT), JSPS Grant-in-Aid for Scientific
Research(A) No. 21244033, and Monbukagakusho 
Grant-in-Aid for Scientific Research on Innovative Area 
Nos. 21111006 and 22111507 are gratefully acknowledged.
This work is also supported by Monbukagakusho
Grant-in-Aid for the global COE program, The Next
Generation of Physics, Spun from Universality and Emergence
at Kyoto University.


\begin{thebibliography}{99}
\bibitem{Hawking:1974sw}
S.~W.~Hawking, Nature {\bf 248} (1974) 30; Commun.\ Math.\ Phys.\  {\bf 43} (1975) 199 [Erratum-ibid.\  {\bf 46} (1976) 206].
\bibitem{Hawking:1980ng}
S.~W.~Hawking, Commun.\ Math.\ Phys.\  {\bf 80}, 421 (1981).
\bibitem{Moss:1984zf}
I.~G.~Moss, Phys.\ Rev.\  D {\bf 32}, 1333 (1985).
\bibitem{Fukushima:2010bq}
K.~Fukushima and T.~Hatsuda, Rept.~Prog.~Phys.~{\bf 74} (2011) 014001
\bibitem{Flachi:2010yz}
A.~Flachi and T.~Tanaka, JHEP {\bf 1102} (2011) 026.
\bibitem{Nickel:2009wj}
D.~Nickel, Phys.\ Rev.\ Lett.\  {\bf 103} (2009) 072301; 
%\bibitem{Nickel:2009wj2}
%D.~Nickel, 
Phys.\ Rev.\  D {\bf 80} (2009) 074025.
\bibitem{Dowker:1989gw}
J.~S.~Dowker, Phys.\ Rev.\  D {\bf 39} (1989) 1235.
\bibitem{Page:1982fm}
D.~N.~Page, Phys.\ Rev.\  D {\bf 25} (1982) 1499.
\bibitem{MacGibbon:2007yq}
J.~H.~MacGibbon, B.~J.~Carr and D.~N.~Page, Phys.\ Rev.\  D {\bf 78}, 064043 (2008); Phys.\ Rev.\  D {\bf 78}, 064044 (2008)
\bibitem{Heckler:1995qq}
A.~F.~Heckler, Phys.\ Rev.\  D {\bf 55} (1997) 480
\bibitem{Fukushima:2003fw}
K.~Fukushima, Phys.\ Lett.\  B {\bf 591} (2004) 277
\end{thebibliography}
\end{document}